# Плавление-кристаллизация наночастиц Sn, Bi и Pb в контакте с Al

**С.И. Богатыренко, С.В. Дукаров, М.М. Колендовский[1], А.П. Крышталь**

*Харьковский национальный университет им. В.Н. Каразина,
пл. Свободы, 4, Харьков, 61077, Украина,
тел. :+3 8(057)70750 47,*

[1]*Научный физико-технологический центр МОН и НАН Украины,
пл. Свободы,6, Харьков, 61077, Украина,*

e-mail address: Sergei.I.Bogatyrenko@univer.kharkov.ua

The results of studies of supercooling upon crystallization value of Bi, Sn and Pb nanosized particles on the Al substrate and between the Al layers have been presented. It has been shown the efficiency of usage of layered film systems for investigation of the limiting supercooling in particle-matrix systems with an eutectic type of interaction between components. The obtained results have been discussed and compared with literature data.

Представлены результаты исследований переохлаждений при кристаллизации наноразмерных частиц Bi, Sn и Pb на Al подложке и между слоями алюминия. Показана эффективность использования слоистых пленочных систем для исследования предельного переохлаждения в системах частица–матрица с эвтектическим типом взаимодействия между компонентами. Полученные результаты обсуждены и сопоставлены с литературными данными.

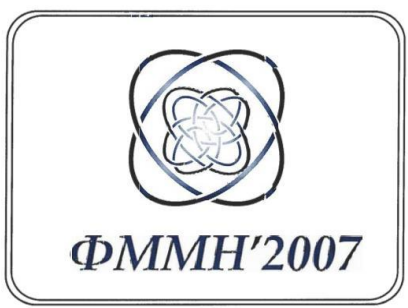

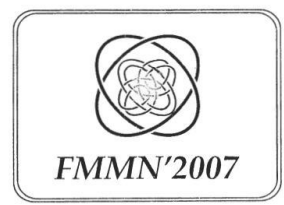

## Плавление-кристаллизация наночастиц Sn, Bi и Pb в контакте с Al

**С.И. Богатыренко, С.В. Дукаров, М.М. Колендовский[1], А.П. Крышталь**

Харьковский национальный университет им. В.Н. Каразина,
пл. Свободы, 4, Харьков, 61077, Украина,
тел.:+38(057)70750 47,
[1]Научный физико-технологический центр МОН и НАН Украины,
пл. Свободы,6, Харьков, 61077, Украина,
e-mail address: Sergei.I.Bogatyrenko@univer.kharkov.ua

*The results of studies of supercooling upon crystallization value of Bi, Sn and Pb nanosized particles on the Al substrate and between the Al layers have been presented. It has been shown the efficiency of usage of layered film systems for investigation of the limiting supercoolings in particle-matrix systems with an eutectic type of interaction between components. The obtained results have been discuses and compared with literature data.*

Развитие наноэлектроники и нанотехнологий в последнее время обратило внимание исследователей к системам с эвтектическим типом взаимодействия, при полной нерастворимости компонентов в твердом состоянии, так как в таких системах возможно совмещение наноразмера частиц с долговечностью и стабильностью их свойств. В плане практического применения перспективны структуры, в которых наночастицы одного компонента, как правило, имеющего меньшую температуру плавления, внедрены в матрицу из более тугоплавкого компонента. Такие системы интенсивно исследуются, однако в литературе встречаются противоречивые данные о температурах фазовых переходов для внедренных частиц. К настоящему времени достоверно установлено, что величина переохлаждения при кристаллизации островковых пленок, полученных в достаточно чистых условиях, зависит от степени их взаимодействия с подложкой, мерой которого является краевой угол смачивания. При этом общая зависимость величины относительного переохлаждения при кристаллизации от краевого угла смачивания металлом подложки для большого количества контактных пар является линейной [1].

В настоящей работе исследованы фазовые переходы в наночастицах легкоплавких металлов Sn, Bi и Pb, находящихся в контакте с алюминием. Выбор объектов обусловлен тем, что указанные системы образуют фазовые диаграммы, которые характеризуются практически полным отсутствием растворимости компонентов в твердом состоянии, при неограниченной растворимости в жидком. Кроме того, в литературе имеется обширное количество данных по величинам переохлаждений при кристаллизации островковых пленок указанных металлов на различных подложках, что предоставляет возможность сравнения и обобщения полученных результатов.

Исследование переохлаждения при кристаллизации островковых пленок Bi, Pb, Sn на алюминиевой подложке проводились с помощью метода, основанного на изучении микроструктуры пленок, конденсированных на подложку с градиентом температуры [2]. Суть метода заключается в том, что на такой подложке визуально наблюдается четкая температурная граница ($T_g$), разделяющая области с различным светорассеянием конденсата, ниже которой пленки практически сплошные, а выше – островковые, причем отдельные островки имеют сферическую форму, что указывает на их конденсацию по механизму пар-жидкость. В результате выполненных экспериментов для исследованных систем были получены значения граничной температуры $T_g$ и краевых углов смачивания легкоплавким металлом подложки, которые приведены в табл. 1.

Следует отметить, что в этом методе величина переохлаждения чувствительна к влиянию неконтролируемых газовых примесей из остаточной атмосферы [1]. Кроме того, определение температуры по смене механизма конденсации для систем с углом смачивания меньше 90° затруднено. Это связанно с незначительной разностью в микроструктуре пленок выше и ниже температуры перехода. На рис.1 показаны электронно-микроскопические снимки пленок

Pb на Al подложке выше и ниже $T_g$. Как можно видеть, достоверное определение границы смены механизма конденсации возможно только при изучении профилей островков.

Таблица 1

| Метал | $\Delta T=T_s-T_g$, К на поверхности Al | $\Delta T=T_s-T_g$, К в Al матрице | Угол смачивания |
|---|---|---|---|
| Bi | 61 | 95 | 75 |
| Sn | 61 | 63 | 42 |
| Pb | 37 | 69 | 50 |

Следует отметить, что в этом методе величина переохлаждения чувствительна к влиянию неконтролируемых газовых примесей из остаточной атмосферы [1]. Кроме того, определение температуры по смене механизма конденсации для систем с углом смачивания меньше 90° затруднено. Это связано с незначительной разностью в микроструктуре пленок выше и ниже температуры перехода. На рис.1 показаны электронно-микроскопические снимки пленок Pb на Al подложке выше и ниже $T_g$. Как можно видеть, достоверное определение границы смены механизма конденсации возможно только при изучении профилей островков.

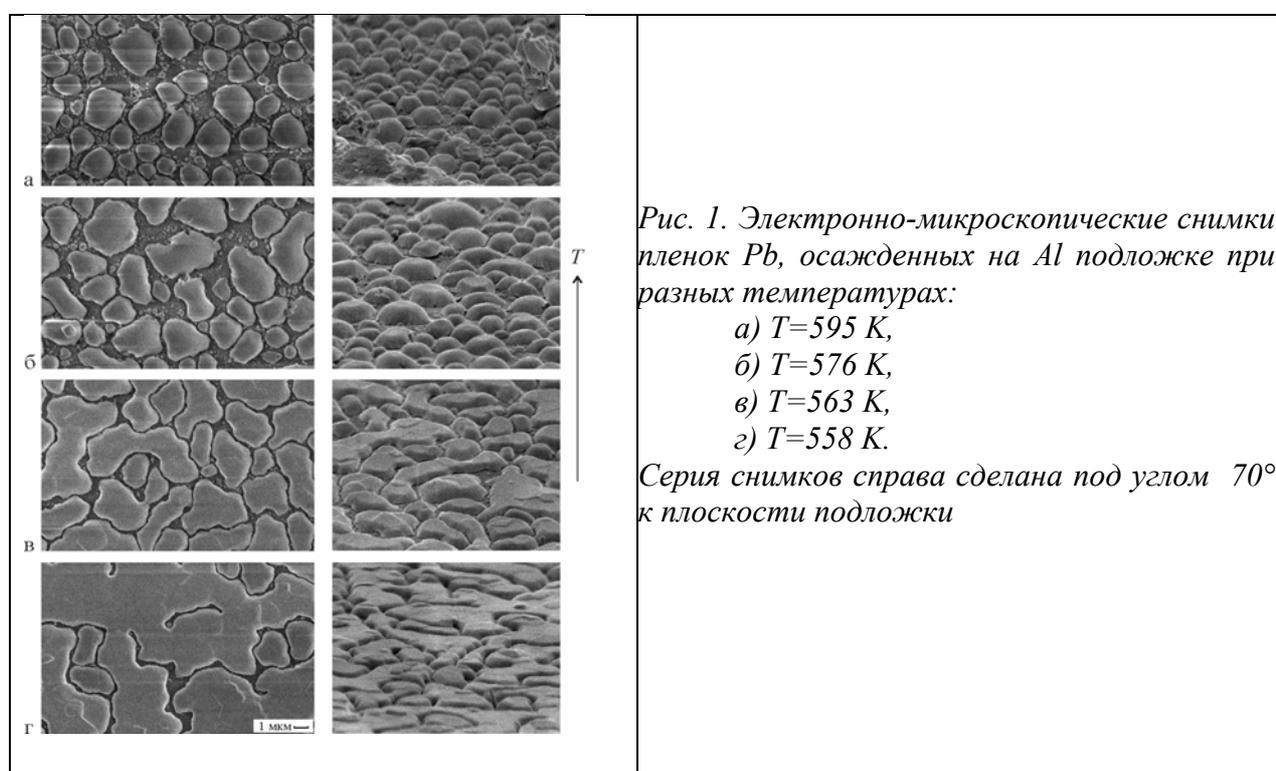

*Рис. 1. Электронно-микроскопические снимки пленок Pb, осажденных на Al подложке при разных температурах:*
  *а) T=595 К,*
  *б) T=576 К,*
  *в) T=563 К,*
  *г) T=558 К.*
*Серия снимков справа сделана под углом 70° к плоскости подложки*

Поэтому в данной работе предложен новый подход, основанный на использовании слоистых пленочных систем, а именно путем измерения их электросопротивления в циклах нагрев-охлаждение. Системы препарировались с большими скоростями путем последовательной вакуумной конденсацией компонентов из независимых источников при комнатной температуре подложки, что позволило минимизировать влияние посторонних твердых и газовых нерастворимых примесей. Единственной примесью, которая и определяет величину переохлаждения – является вещество матрицы, в данном случае Al.

Схема эксперимента была следующей. На стеклянную пластину размером 30×20 мм в вакууме путем термического испарения наносились контактные дорожки из алюминия. Подложка с контактами крепилась к медному блоку-нагревателю с соответствующей маской. Система расположения контактных дорожек, показанная на рис.2, позволяла в одном эксперименте одновременно в совершенно идентичных условиях проводить измерения электросопротивления в трех слоистых пленочных системах.

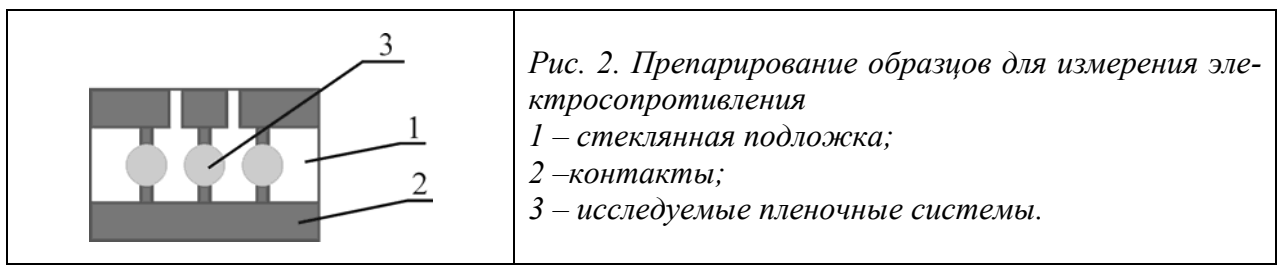

*Рис. 2. Препарирование образцов для измерения электросопротивления*
*1 – стеклянная подложка;*
*2 – контакты;*
*3 – исследуемые пленочные системы.*

Эксперименты проводились в вакуумной установке с безмасляной системой откачки при давлении остаточных газов ниже $1\times10^{-7}$ мм рт.ст. Нагрев и охлаждение осуществлялись в интервале от комнатной температуры и приблизительно до 670 K.

Оказалось, что электросопротивление образцов очень чувствительно к процессам, происходящим в пленках при их нагреве и охлаждении. Так в слоистых пленочных системах Al/M/Al (M = Sn, Bi, Pb) обнаружен температурный гистерезис плавление-кристаллизация для фазы на основе легкоплавкого компонента. Так, на рис.3 показана типичная зависимость сопротивления для системы Al/Bi/Al от температуры при нагреве и охлаждении. Видно, что непосредственно после конденсации в процессе нагревания сопротивление при повышении температуры плавно возрастает, а при приближении к температуре плавления эвтектики на основе висмута – падает подобно электросопротивлению чистого висмута. Выше этой температуры сопротивление слабо возрастает, что соответствует случаю сплава. При охлаж-днении сопротивление плавно уменьшается до температуры соответствующей кристалли-зации сплава, при которой наблюдается скачкообразное увеличение его почти в 1,5 раза.

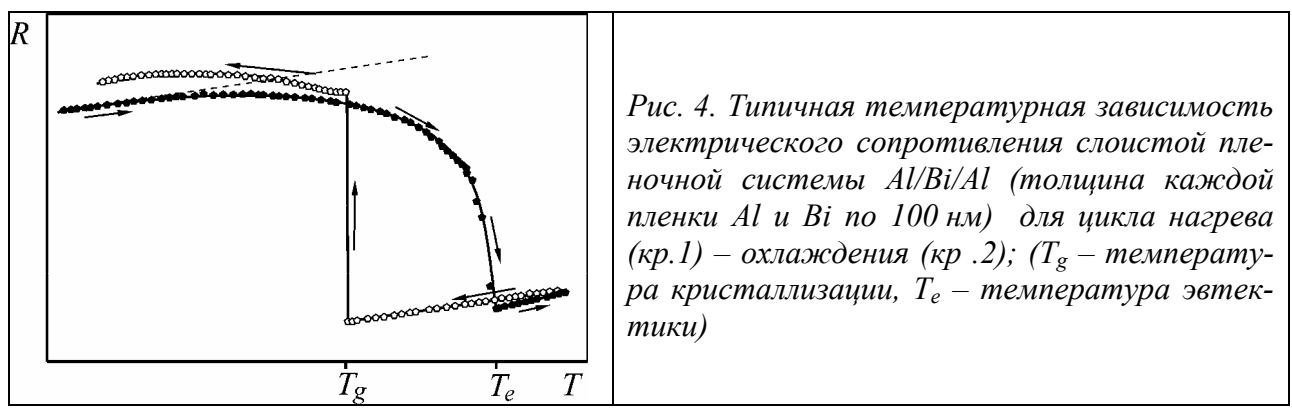

*Рис. 4. Типичная температурная зависимость электрического сопротивления слоистой пленочной системы Al/Bi/Al (толщина каждой пленки Al и Bi по 100 нм) для цикла нагрева (кр.1) – охлаждения (кр .2); ($T_g$ – температура кристаллизации, $T_e$ – температура эвтектики)*

Аналогичные зависимости были получены и для систем на основе олова и свинца, что указывает на то, что такое поведение электросопротивления связано именно с характером взаимодействия компонент, и вероятно, будет иметь место для всех систем эвтектического типа. Данные о величинах переохлаждения частиц исследованных металлов между слоями алюминия представлены в таблице 1. Видно, что использование слоистых пленочных систем Al/Bi/Al и Al/Pb/Al позволяет достичь более глубоких переохлаждений по сравнению с другими методиками, поскольку указанные системы наиболее чувствительны к влиянию примесей из остаточных газов.

Полученные результаты хорошо согласуются с общей зависимостью величины относительного переохлаждения при кристаллизации от краевого угла смачивания металлом подложки [2].

**Литература**